\documentclass[10pt]{article}
\usepackage{fullpage}
\usepackage{amsmath}
\usepackage{amssymb}
\usepackage[dvips]{epsfig}

\usepackage{graphicx}
\usepackage{color}
\usepackage{cite}
\usepackage{upgreek}

\def\blue{\textcolor{black}}

\def\##1{\underline{#1}}
\def\=#1{\underline{\underline{#1}}}

\def\+#1{\underline{\bf #1}}
\def\*#1{\underline{\underline{\bf #1}}}

\def\.{\mbox{ \tiny{$^\bullet$} }}

\def\eps{\varepsilon}

\def\lambdao{\lambda_{\scriptscriptstyle 0}}

\newcommand{\SImu}{\ensuremath{\upmu}}
\newcommand{\SImum}{\SImu\textrm{m}}

\def\ux{\#{u}_x}
\def\uy{\#{u}_y}
\def\uz{\#{u}_z}

\def\un{\#{u}_n}
\def\ut{\#{u}_\tau}
\def\ub{\#{u}_b}

\def\c#1{\cite{#1}}
\def\l#1{\label{#1}}
\def\r#1{(\ref{#1})}

\def\le{\left(}
\def\ri{\right)}
\def\les{\left[}
\def\ris{\right]}
\def\lec{\left\{}
\def\ric{\right\}}

\def\refo{^{\rm o}_{\rm ref}}

\begin{document}

\begin{center}

{\bf {\Large Polarization-state-dependent attenuation and amplification in a columnar thin film}
 }

 \vspace{5mm} \large

  \large

Tom G. Mackay\footnote{University of Edinburgh, School of Mathematics and Maxwell Institute for Mathematical Sciences, 
Edinburgh EH9 3FD, UK; Pennsylvania State University, Department of Engineering Science and Mechanics,  University Park, PA 16802, USA; email:  T.Mackay@ed.ac.uk}
 and
Akhlesh Lakhtakia\footnote{Pennsylvania State University, Department of Engineering Science and Mechanics,  University Park, PA 16802, USA}

\vspace{2mm}

\vspace{2mm}

\normalsize

\vspace{2mm}

\end{center}

\begin{abstract}
We numerically investigated the plane-wave reflection--transmission characteristics of 
a columnar thin film (CTF) whose columns are made from a dissipative material but whose void regions are filled with an active material.
By computing the  reflectances and transmittances, we found that the CTF can simultaneously amplify $s$-polarized incident light and attenuate 
$p$-polarized incident light, or vice versa. This polarization-state-dependent  attenuation and amplification phenomenon depends upon the angle of incidence  and the thickness of the CTF.

\end{abstract}

\vspace{2mm}

\vspace{2mm}

The advent of nanotechnology and the emergence of metamaterials \c{Walser} has allowed composite materials with exotic optical properties to be realistically 
contemplated and, in some cases, actually realized.
In particular,  theoretical studies have revealed attractive response characteristics associated with composite materials containing both active and dissipative component materials. While active component materials have been widely incorporated into various metamaterials in an attempt to combat intrinsic losses \c{Sun_APL,Wuestner,Dong_APL,Strangi}, other applications of active component materials have cropped up more recently.
For example, the homogenization of a mixture of both active and dissipative spheroidal particles engenders a uniaxial dielectric material    that simultaneously exhibits  both amplification and attenuation, depending upon the direction of propagation \c{ML_PRA}. 
In a similar vein, interleaving layers of  active and dissipative component materials leads to a birefringent material which facilitates arbitrary control over polarization states \c{Fan_PRL}.
Homogenization can also lead to an isotropic chiral material that amplifies  left-circularly polarized light but attenuates  right-circularly polarized light (or vice versa) \c{ML_chiral}. In this Letter we report on the ability of a columnar thin film (CTF) comprising both dissipative and active component materials to discriminate between incident light of $s$ and $p$ polarization states.

A CTF
is an array of parallel identical  nanocolumns which
may be fabricated  \blue{by
oblique-angle physical-vapor-deposition techniques that include thermal evaporation, electron-beam
evaporation, sputtering, and pulsed-laser deposition}
  \c{HW,STF_Book}.
Macroscopically, the optical properties of CTFs are equivalent to those of certain orthorhombic crystals.
CTFs are attractive
platforms for
applications such as
 sensing chemical and biological species
\c{LMBR,PH07}, due to
the ability to   
engineer their macroscopic optical responses
via  control over
the porosity and the columnar
morphology   at the fabrication stage.

Consider a CTF with relative permittivity dyadic
\begin{equation}
\=\eps = \=S_{\,y} (\chi) \.  \=\eps\refo \. \=S^T_{\,y} (\chi),
\end{equation}
where the diagonal dyadic
\begin{equation}
 \=\eps\refo = \eps_a \, \uz \, \uz + \eps_b \, \ux \, \ux + \eps_c \,\uy \, \uy \, ,
\end{equation}
signifies macroscopic  orthorhombic symmetry \c{EAB},
while the rotation dyadic
\begin{eqnarray}
 \=S_{\,y} (\chi) & = & \uy \, \uy +  \le \ux \, \ux + \uz \, \uz \ri \cos \chi \nonumber \\ && + \le  \uz \, \ux - \ux \, \uz \ri \sin \chi
\end{eqnarray}
involves the nanocolumn inclination angle $\chi \in(0,\pi/2]$. The unit vectors aligned with the Cartesian axes are represented by the triad $\lec \ux, \uy, \uz \ric$.
  The CTF  occupies the region between $z=0$ and $z=L$.
  
\begin{figure}
\begin{center}
\begin{tabular}{c}
\includegraphics[height=5.5cm]{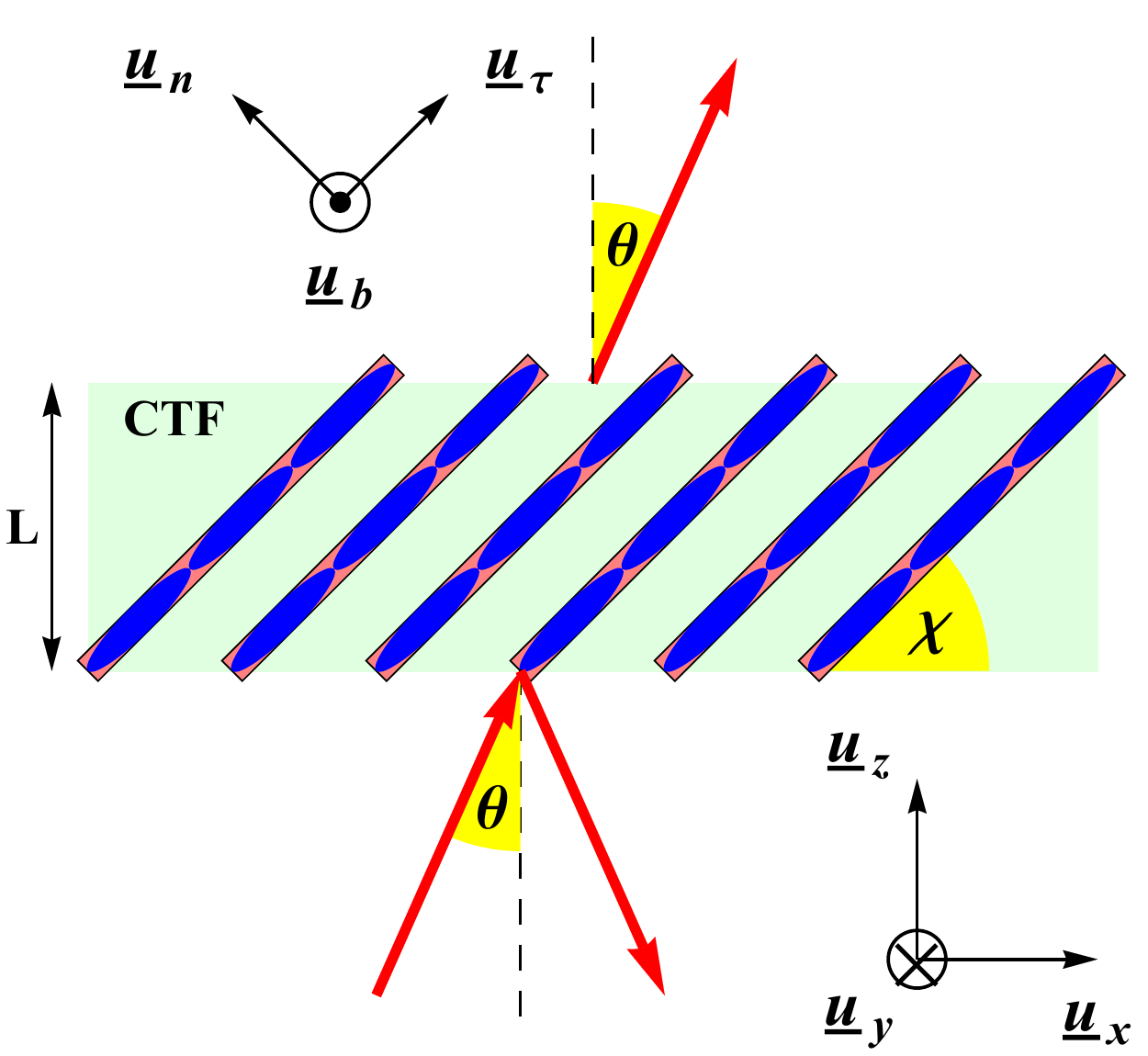}
\end{tabular}
\end{center}
\caption { \label{fig1}  A schematic representation of  reflection--transmission for a CTF.
} 
\end{figure}

Numerical values for the three relative-permittivity parameters $\eps_{a} $, $\eps_{b} $, and $\eps_{c} $ can be estimated using a standard homogenization procedure \c{STF_Book,SLH}. To this end,
each column of the CTF may be viewed, at the nanoscopic scale, as a collection of highly elongated
ellipsoidal inclusions strung together end-to-end, with  all inclusions having the same orientation and shape.
The  dyadic
\begin{equation}
 \un \, \un + \gamma_\tau \, \ut \, \ut + \gamma_b \, \ub \, \ub
\end{equation}
specifies the shape of each ellipsoid. Herein
 the normal, tangential, and binormal basis vectors are specified
in terms of the nanocolumn inclination angle $\chi$  per
\begin{equation}
\left. \begin{array}{l}
 \un = - \ux \, \sin \chi + \uz \, \cos \chi \vspace{4pt} \\
 \ut =  \ux \, \cos \chi + \uz \, \sin \chi \vspace{4pt} \\
\ub = - \uy
\end{array}
\right\},
\end{equation}
whereas $\gamma_b>0$ and $\gamma_\tau>0$ are shape factors.
The CTF is represented  schematically in Fig.~\ref{fig1}.

Since the nanocolumnar morphology is highly aciculate, 
we fixed $\gamma_\tau = 15$ and  $\gamma_{b} = 2$, in keeping with earlier studies \c{ML_JNP}.
Also  the   nanocolumns were taken to be  made from magnesium monoxide  \cite{SM} impregnated with  about  0.2\% v/v silver nanoparticles.  Thus, the refractive index of the columnar material was  estimated as   $1.7310+ 0.0086 i$
using the Biot--Arago formula \cite{BA1806}, when the free-space wavelength $\lambdao=637$~nm.
The void regions  of the CTF were supposed to be filled with an active material which is taken to be a mixture of rhodamine 800 and rhodamine 6G. Such a mixture can  possess a relative permittivity with imaginary part in the range
$\le -0.15, -0.02 \ri$ and real part in the range $\le 1.8, 2.3 \ri$ across the 
440--500-THz frequency range,
depending upon the relative concentrations  and the external pumping rate \c{Sun_APL}.
We took
 the relative permittivity of the mixture to be $2 - 0.02 i$ for our calculations. 
 
\begin{figure}
\begin{center}
\begin{tabular}{c}
\includegraphics[scale=0.6]{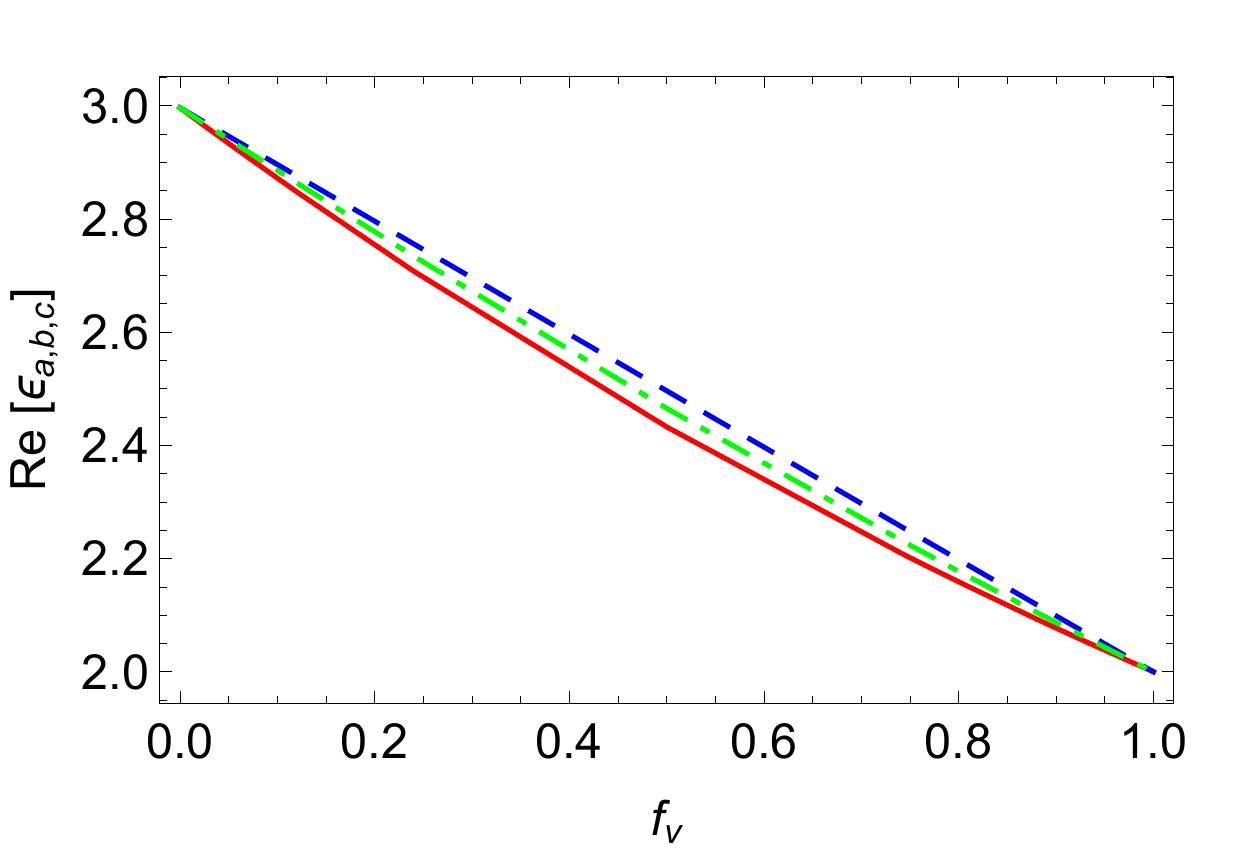}\\
\includegraphics[scale=0.6]{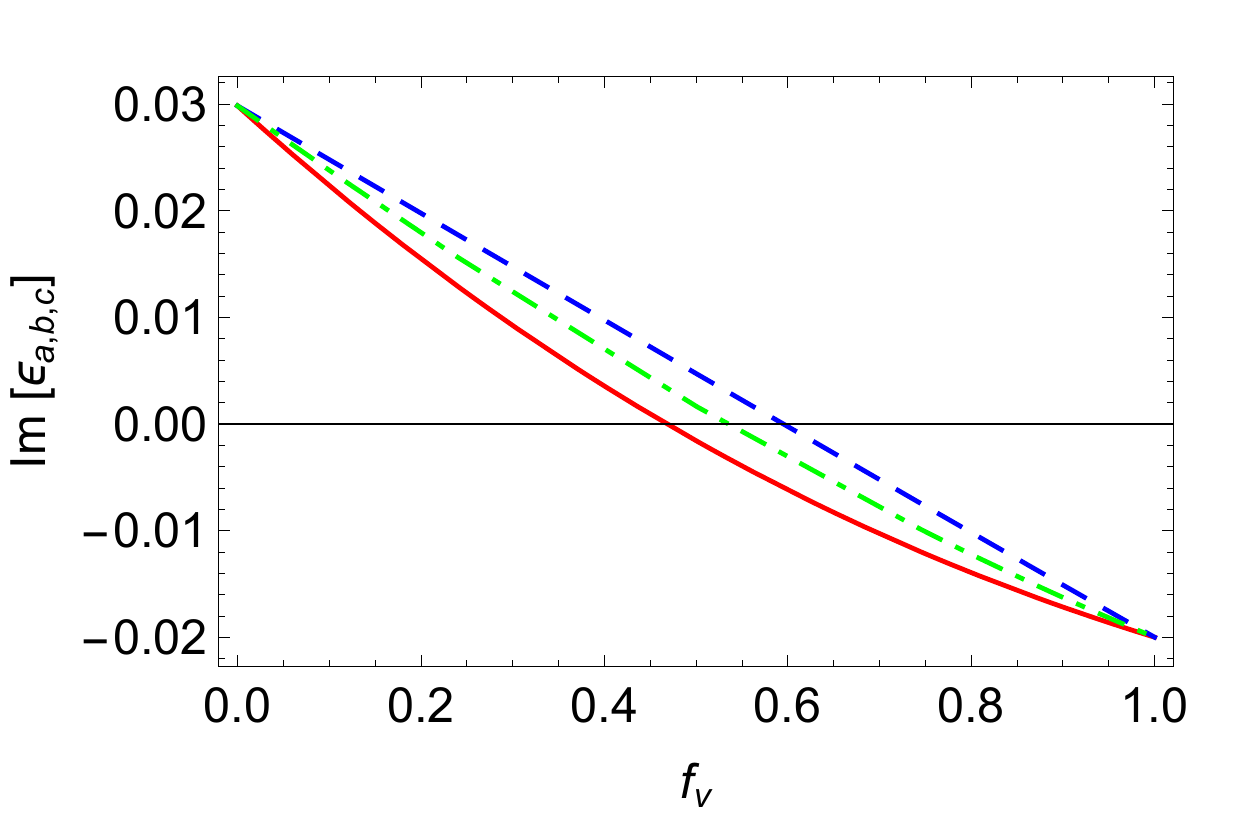}
 \end{tabular}
\end{center}
 \caption{\label{fig2}
 Real and imaginary parts of the relative permittivity parameters $\eps_a$ (red solid curves), $\eps_b$ (blue dashed  curves), and
 $\eps_c$ (green broken dashed curves) plotted against porosity $f_v$.
 } 
\end{figure}

 Following the methodology described elsewhere \c{ML_JNP},  $\eps_{a} $, $\eps_{b} $, and $\eps_{c} $ were estimated using the Bruggeman homogenization formalism \c{MAEH}. These relative permittivity parameters  are plotted against porosity $f_v \in \le 0,1 \ri$ in Fig.~\ref{fig2}. For a narrow range of porosity  $f_v \in \le 0.46, 0.60 \ri $ 
 the imaginary parts of $\eps_{a,b,c}$ have different signs, which indicates that the CTF simultaneously allows   attenuation and amplification  \c{ML_PRA}.
For the remainder of this Letter, we set
 $f_v= 0.53$ at which $\eps_a = 2.4033 - 0.0030i$,  $\eps_b = 2.4665 + 0.0032i$,  and $\eps_c = 2.43570 + 0.0002i$.
Also, the nanocolumn inclination angle was fixed at $\chi = 60^\circ$.

Now we turn to the reflection--transmission problem for the CTF. Suppose that an incident plane wave exists in the vacuous  half-space $z<0$, as represented by the electric field phasor
\begin{eqnarray}
\nonumber
&&\#E_{\rm inc} = \le a_s \#s + a_p \#p_{\,+} \ri 
\exp\left\{i (2\pi/\lambdao) \right.
\\
&&\qquad\times
\left.
\les  \le x \cos \psi + y \sin \psi \ri \sin \theta +  z \cos \theta \ris\right\},
\end{eqnarray}
decomposed into $s$-  and $p$-polarized components, with
\begin{equation}
\left.
\begin{array}{l}
\#s = - \ux \sin \psi + \uy \cos \psi \\
\#p_{\, \pm} = \mp \le \ux \cos \psi + \uy \sin \psi \ri \cos \theta + \uz \sin \theta 
\end{array}
\right\}.
\end{equation}
\blue{Whereas
$\theta$ is the angle of incidence with respect to the $z$ axis, $\psi$ is the angle of incidence with
respect to the $x$ axis in the $xy$ plane.}
 Correspondingly, the reflected plane wave represented by
\begin{eqnarray}
\nonumber
&&\#E_{\rm ref} = \le r_s \#s + r_p \#p_{\,-} \ri 
\exp\left\{i (2\pi/\lambdao) \right.
\\
&&\qquad\times
\left. \les  \le x \cos \psi + y \sin \psi \ri \sin \theta -  z \cos \theta \ris\right\}
\end{eqnarray}
also exists in  the half-space $z<0$, and the transmitted plane wave represented by
\begin{eqnarray}
\nonumber
&&\#E_{\rm tr} = \le t_s \#s + t_p \#p_{\,+} \ri 
\exp\left\{i (2\pi/\lambdao) \right.
\\
&&\qquad\times
\left. \les  \le x \cos \psi + y \sin \psi \ri \sin \theta + \le z -L \ri  \cos \theta \ris\right\}
\end{eqnarray}
exists in the vacuous half-space $z>L$. 
 
Following  a standard procedure \c{STF_Book}, the two  reflection
amplitudes $r_{s,p}$ and the two transmission
amplitudes $t_{s,p}$ may be found in terms of the two
incidence amplitudes $a_{s,p}$ per
\begin{equation}
\les 
\begin{array}{c}
r_s \\
r_p
\end{array}
\ris =
\les 
\begin{array}{cc}
r_{ss} & r_{sp} \\
r_{ps} & r_{pp}
\end{array}
\ris
 \les 
\begin{array}{c}
a_s \\
a_p
\end{array}
\ris
\end{equation}
and
\begin{equation}
\les 
\begin{array}{c}
t_s \\
t_p
\end{array}
\ris =
\les 
\begin{array}{cc}
t_{ss} & t_{sp} \\
t_{ps} & t_{pp}
\end{array}
\ris
 \les 
\begin{array}{c}
a_s \\
a_p
\end{array}
\ris,
\end{equation}
wherein the four reflection coefficients $r_{ss,sp,ps,pp}$ and 
the four transmission coefficients 
$t_{ss,sp,ps,pp}$  have been introduced. The corresponding reflectances are defined as $R_{sp} = |
r_{sp} |^2$ etc., and the corresponding transmittances defined as $T_{ps} = |
t_{ps} |^2$ etc.

For a wholly dissipative material (i.e., one that does not simultaneously exhibit attenuation and amplification), it follows from the principle of 
energy conservation that
\begin{equation} \l{inequal}
\left.
\begin{array}{l}
R_{ss} + R_{ps} + T_{ss} + T_{ps} < 1 \\
R_{pp} + R_{sp} + T_{pp} + T_{sp} < 1
\end{array}
\right\}.
\end{equation}
Similarly, for a wholly active material the inequalities \r{inequal} are reversed, i.e.,
\begin{equation} \l{inequal2}
\left.
\begin{array}{l}
R_{ss} + R_{ps} + T_{ss} + T_{ps} > 1 \\
R_{pp} + R_{sp} + T_{pp} + T_{sp} > 1
\end{array}
\right\}.
\end{equation}

\begin{figure}
\begin{center}
\begin{tabular}{c}
\includegraphics[scale=0.6]{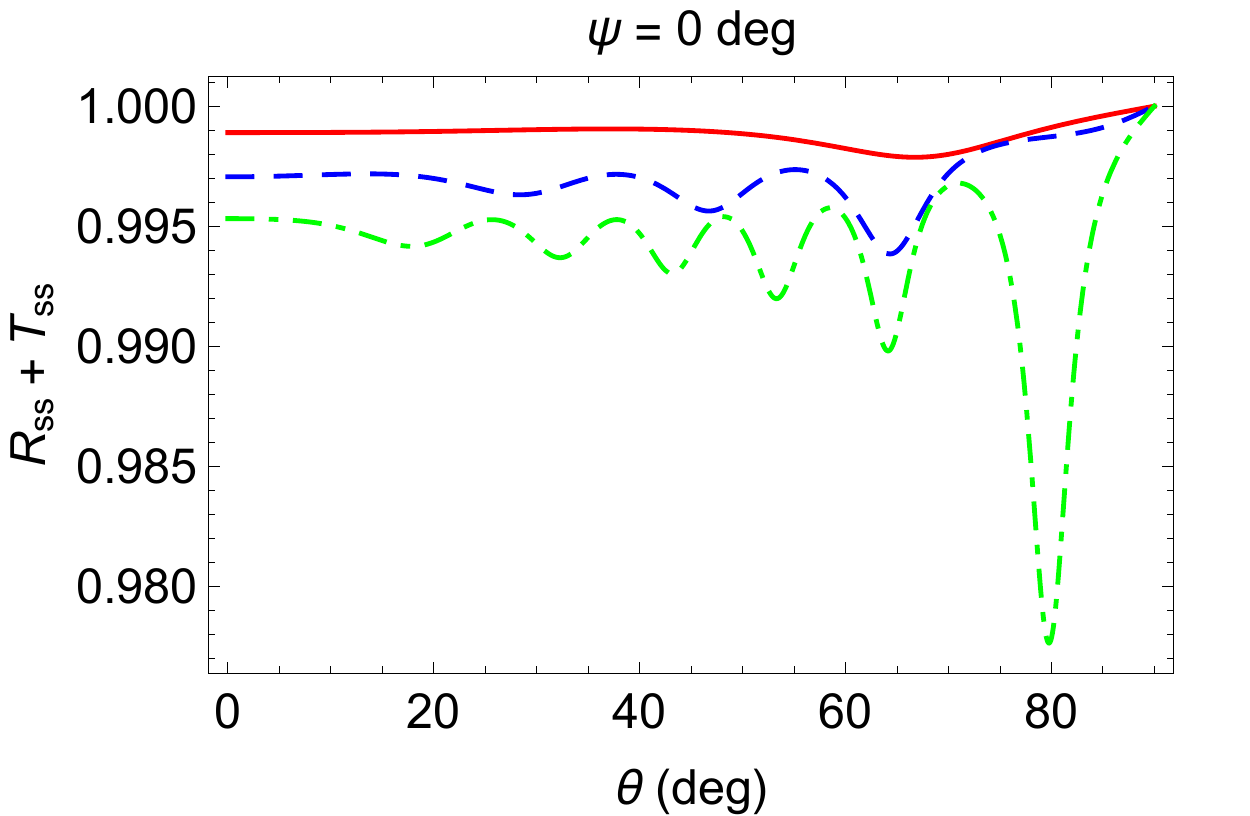}\\
\includegraphics[scale=0.6]{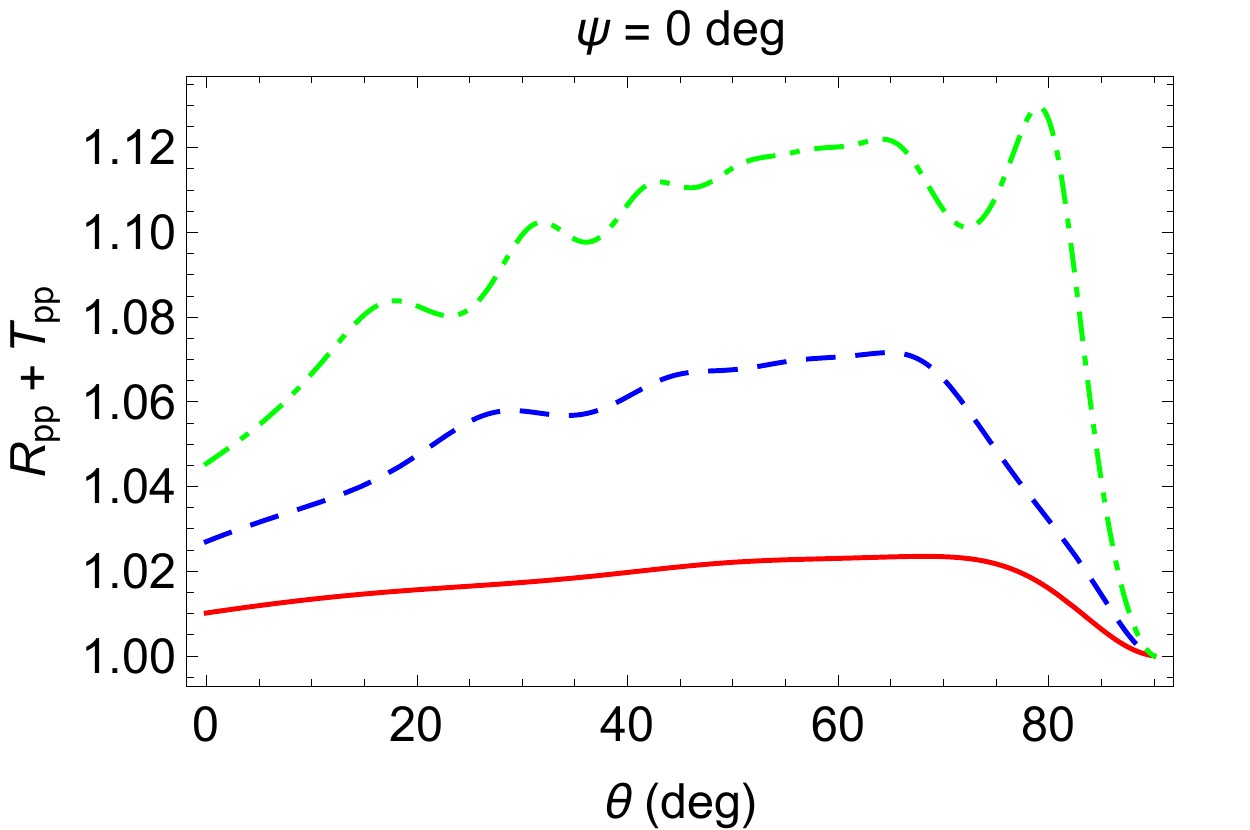}
 \end{tabular}
\end{center}
 \caption{\label{fig3}  The reflectance--transmittance   sums $R_{ss}+T_{ss} $ and $R_{pp}+T_{pp} $  plotted against $\theta$
for $\psi = 0^\circ$, when   $L  = 1$~\SImum  ~(red solid curves),  $ 3$~\SImum   ~(blue dashed  curves), and $ 5$~\SImum ~(green broken dashed curves).
 } 
\end{figure}

All eight remittances  were computed for $\lambdao = 637$ nm.  When the azimuthal  angle $\psi = 0^\circ$,  all four cross-polarized remittances, namely $R_{sp, ps}$ and 
$T_{sp, ps}$,  are null valued.   For $\psi = 0^\circ $,
the remittance sum $R_{ss}+T_{ss} $ for $s$-polarized incident light and 
the remittance sum $R_{pp}+T_{pp} $ for $p$-polarized incident light are plotted against the polar angle $\theta$ for the CTF
 thicknesses  $L \in \lec 1 ,  3,  5 \ric$~\SImum
~in Fig.~\ref{fig3}. For all values  of the polar angle $\theta$, we see that  $R_{ss}+T_{ss} < 1 $ whereas   $R_{pp}+T_{pp} >1  $. Thus, $p$-polarized incident light is amplified whereas
$s$-polarized incident light is attenuated inside the CTF.

This difference in the responses to $s$-polarized incident light and $p$-polarized incident light is generally enhanced as the CTF thickness increases. That is, the degree of amplification for $p$-polarized incident light is greater for larger values of $L$ and the degree of attenuation for $s$-polarized incident light is greater for larger values of $L$.

\begin{figure}
\begin{center}
\begin{tabular}{c}
\includegraphics[scale=0.6]{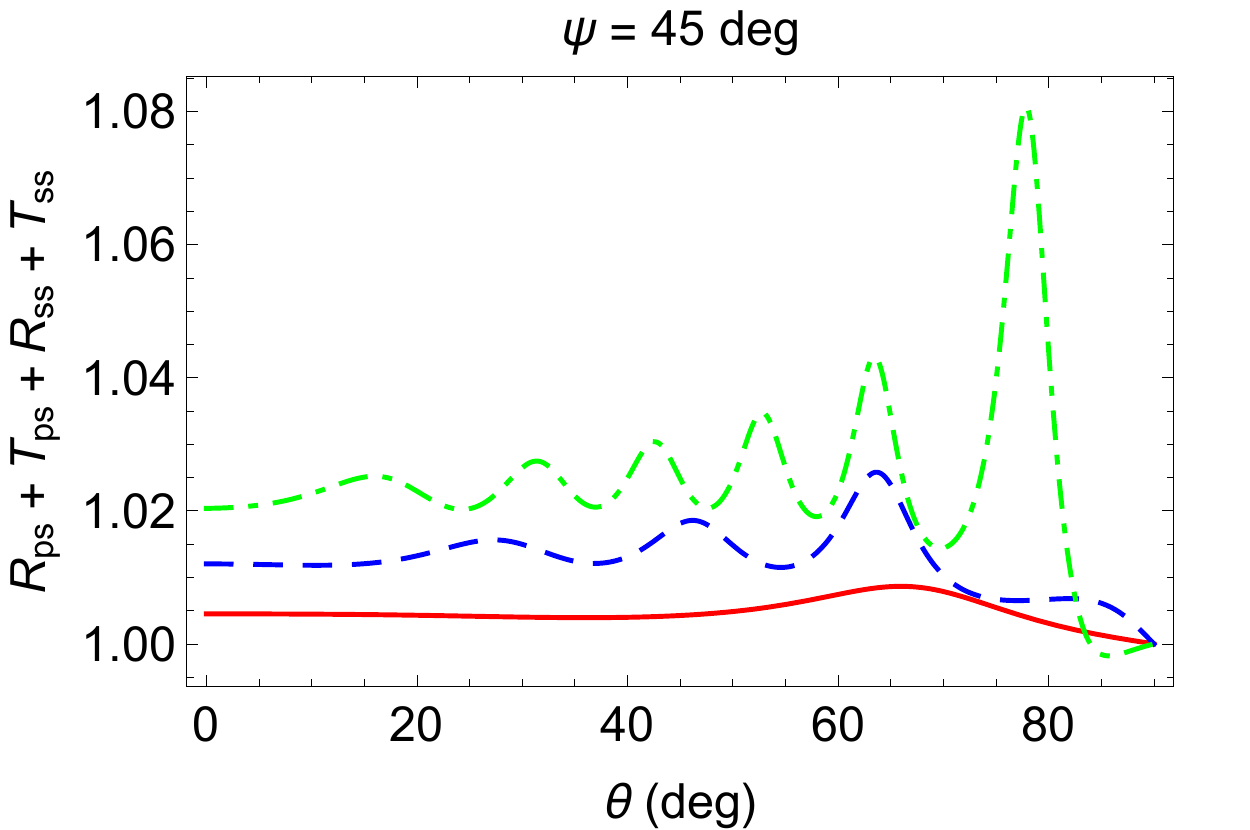}\\
\includegraphics[scale=0.6]{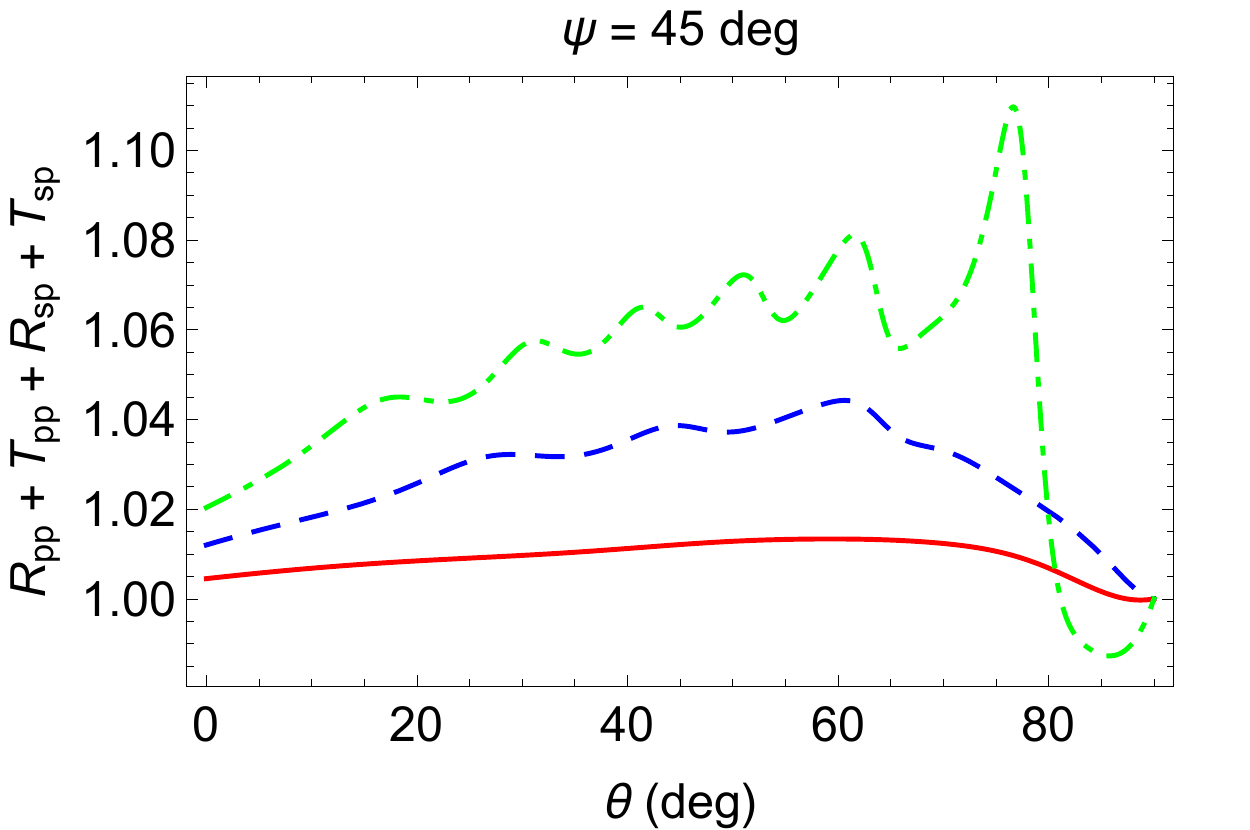}
 \end{tabular}
\end{center}
 \caption{\label{fig4}  The reflectance--transmittance   sums  $R_{ss}+R_{ps}+T_{ss}+T_{ps} $ and $R_{pp}+R_{sp}+T_{pp}+T_{sp} $  plotted against $\theta$  
for  $\psi = 45^\circ$,  when $L  = 1$~\SImum  ~(red solid curves),  $3$~\SImum   ~(blue dashed  curves), and $5$~\SImum  ~(green broken dashed curves).
 } 
\end{figure}

For   $\psi = 45^\circ $,
the remittance sum $R_{ss} + R_{ps} + T_{ss} + T_{ps} $ for $s$-polarized incident light and 
the remittance sum $R_{pp} + R_{sp} + T_{pp} + T_{sp} $ for $p$-polarized incident light are plotted against the polar angle $\theta$ for thicknesses  $L \in \lec 1 , 3, 5\ric$~\SImum 
~in Fig.~\ref{fig4}. 
The picture  here is quite different to that for  $\psi = 0^\circ$. When $\psi = 45^\circ$ and $\theta< 81^\circ$, the
    incident light is  amplified  regardless of its polarization state, and the degree of amplification generally increases as the  CTF thickness increases.
However,  $p$-polarized incident light is attenuated for $L = 5$~\SImum~but amplified for smaller values of $L$ when $\theta > 81^\circ$.

\begin{figure}
\begin{center}
\begin{tabular}{c}
\includegraphics[scale=0.6]{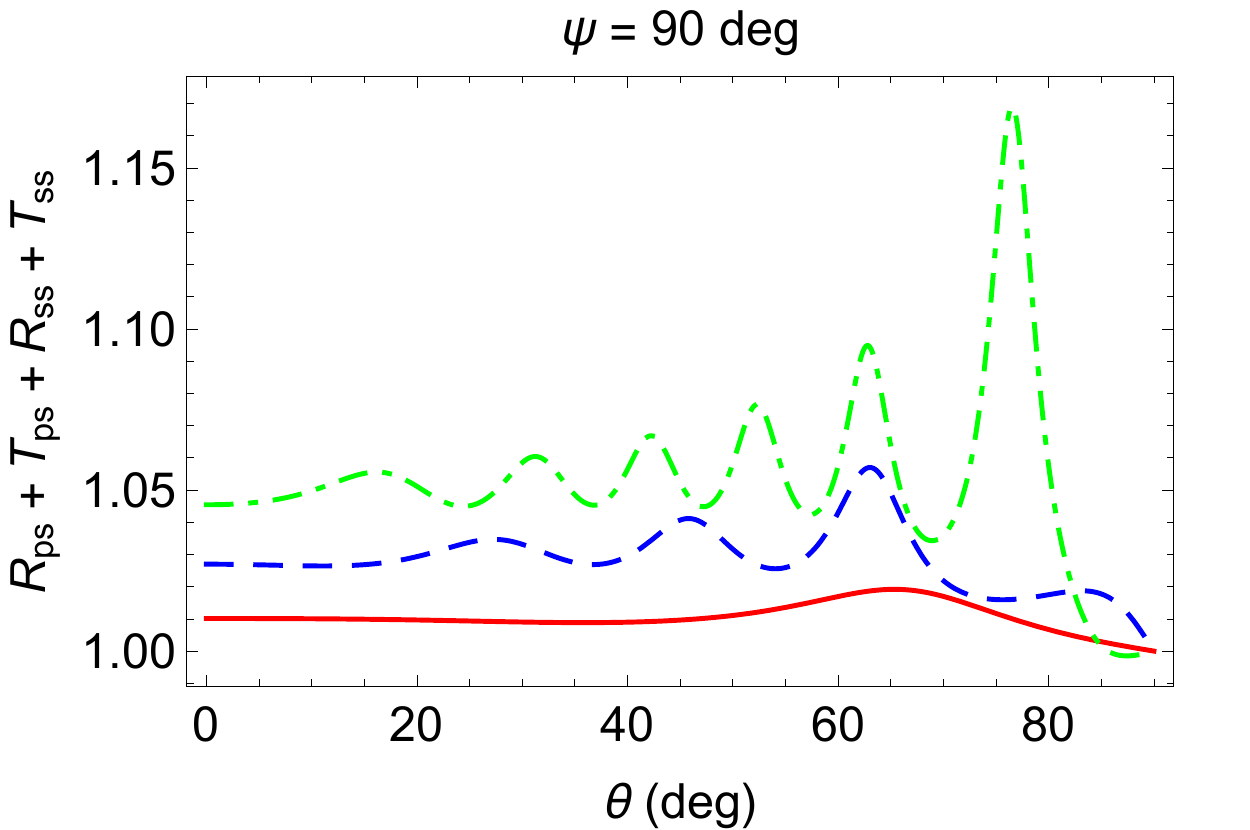}\\
\includegraphics[scale=0.6]{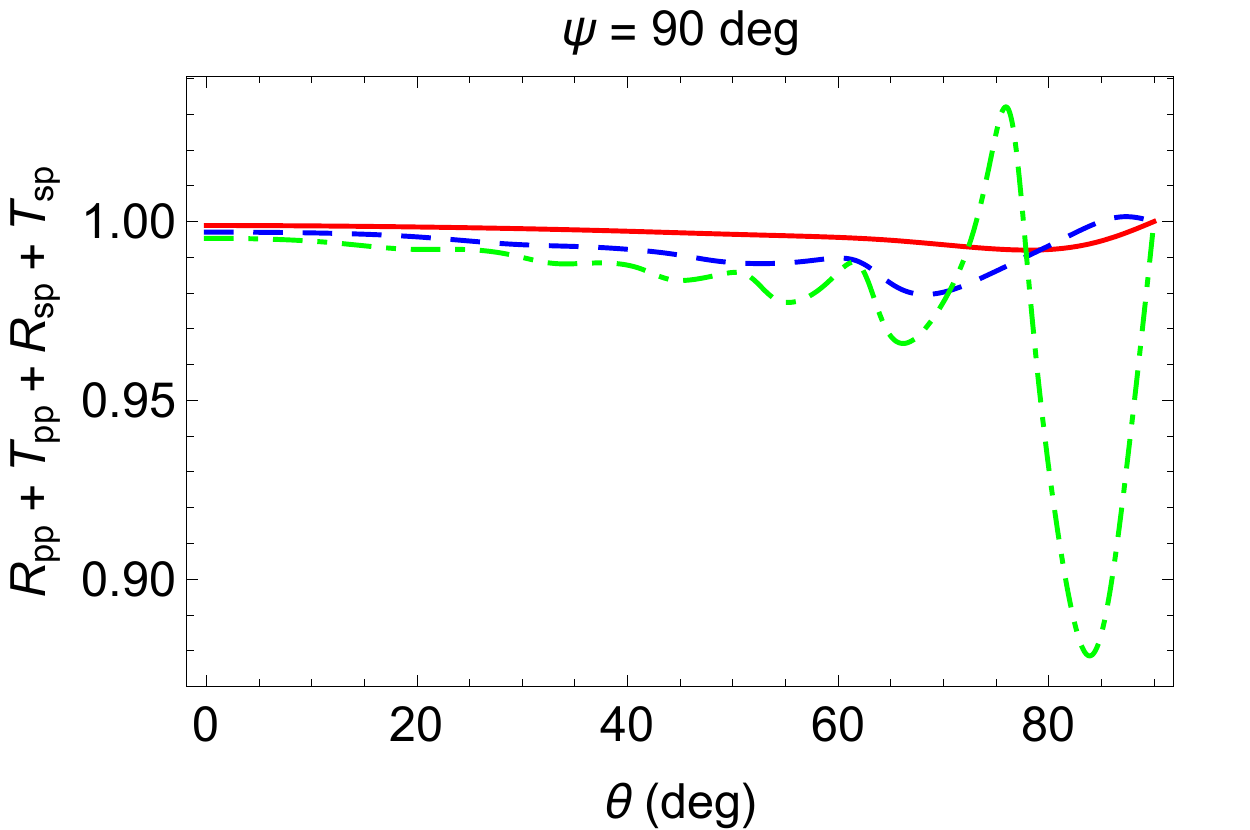}
 \end{tabular}
\end{center}
 \caption{\label{fig5} As Fig.~\ref{fig4} except that $\psi = 90^\circ$.
 } 
\end{figure}

Lastly, we turn to the case where $\psi = 90^\circ $, as represented in Fig.~\ref{fig5} wherein the  
remittance sums   $R_{ss} + R_{ps} + T_{ss} + T_{ps} $ and $R_{pp} + R_{sp} + T_{pp} + T_{sp} $ are plotted against  $\theta$ for  $L \in \lec 1 , 3, 5\ric$~\SImum. In contrast to what happens for $\psi = 0^\circ $, here  $s$-polarized incident light is generally amplified while 
 $p$-polarized incident light is generally attenuated. As in Figs.~\ref{fig4} and \ref{fig5}, the degrees of amplification and attenuation are generally exaggerated as the CTF thickness increases.
However, there are notable exceptions. For example, for a narrow range of polar angles centered on $\theta = 76^\circ$, $p$-polarized incident  light is amplified for $L= 5$~\SImum  ~but attenuated for smaller values of $L$.

The CTF considered  in this Letter  represents a practicable proposition, with realistic values being chosen for the relative permittivities of the component materials.
By combining a dissipative  material for the nanocolumns with an active material to fill the void regions of a CTF, polarization-state-dependent attenuation and amplification may be achieved. \blue{Parenthetically, the concept of polarization--state--dependent attenuation and amplification has previously been described for a nonspecific optical system \c{Azzam}.  In contrast, the 
phenomenon described herein concerns a physically-realizable engineered material. }
 These results open the door for linear polarizers of a
new type.

\vspace{5mm}

\noindent {\bf Acknowledgement}  AL is grateful to the Charles Godfrey
Binder Endowment at Penn State for ongoing support of his research.

\end{document}